\documentclass{PoS}
\usepackage{graphicx}
\usepackage{nicefrac}
\usepackage[modulo]{lineno}
\usepackage{amsmath}

\title{A method to reconstruct the muon lateral distribution with an array of segmented counters with time resolution}

\ShortTitle{Muon reconstruction}

\author{ D. Ravignani$\,^a$,  A. D. Supanitsky$\,^b$, D. Melo$\,^a$, and \speaker{B. Wundheiler}$^{,\,a}$ \\ 
\llap{$^a$} ITeDA (CNEA, CONICET, UNSAM), Buenos Aires, Argentina \\
\llap{$^b$} IAFE (CONICET, UBA), Buenos Aires, Argentina \\
E-mail: \email{diego.ravignani@iteda.cnea.gov.ar} }

\abstract{Despite the significant experimental effort made in the last decades, the origin of the ultra high energy cosmic rays is 
still unknown. The chemical composition of these energetic particles carries key astrophysical information 
to identify where they come from. It is well known that the muon content of the showers generated
by the interaction of the cosmic rays with air molecules, is very sensitive to the primary
particle type. Therefore, the measurement of the number of muons at ground level is an essential tool to infer the cosmic
ray mass composition. We introduce a novel method to reconstruct the lateral distribution of muons
with an array of counters buried underground like AMIGA, one of the Pierre Auger  
Observatory detector systems. The reconstruction builds on a previous method we recently presented by considering the detector 
time resolution. With the new method more events can be reconstructed than with the previous one.
In addition the statistical uncertainty of the measured number of muons is reduced, allowing for a better primary mass discrimination.}

\FullConference{The 34th International Cosmic Ray Conference,\\
		30 July- 6 August, 2015\\
		The Hague, The Netherlands}

\begin{document}


\section{Introduction}

Although the origin of the ultra high energy cosmic rays is still unknown, significant progress has been recently achieved  
from data collected by cosmic rays observatories like the Pierre Auger Observatory~\cite{Aab:2015zoa} and the Telescope Array~\cite{AbuZayyad:2012kk}. The three main observables used to study the nature of cosmic rays are their energy spectrum, arrival directions, and chemical composition. 
Certainly composition is a crucial ingredient to understand the origin of these very energetic particles~\cite{Kampert:2012mx}, to find the spectral region where the 
transition between the galactic and extragalactic cosmic rays takes place~\cite{MedinaTanco:2007gc},
and to elucidate the origin of the flux suppression at the highest energies~\cite{Kampert:2013dxa}.

For energies larger than $10^{15} \,\mathrm{eV}$ the cosmic rays are studied by observing the atmospheric showers produced
when they interact with the air molecules. Therefore composition has to be inferred indirectly by using parameters measured in air shower observations. The parameters most sensitive 
to the primary mass are the depth of the shower maximum, and the number of muons generated during the cascade process. 
While the maximum depth is obtained from fluorescence telescopes, the number of muons at ground level is measured with dedicated muon detectors.
At the highest cosmic ray energies the hadronic interactions are unknown, so models that extrapolate accelerator data at lower centre of mass energy are used in shower simulations. As the number of muons predicted by simulations strongly depends on the 
assumed high energy hadronic interaction model, the measurement of the muon component is also a very important tool to understand the hadronic interactions at cosmic ray energies~\cite{Supanitsky:2008ph}.

AMIGA is a detector system of the Auger Observatory being constructed that includes a triangular array of muon counters spaced at $750\,\mathrm{m}$~\cite{Wundheiler:2015}. 
Each array position has three $10\,\mathrm{m^2}$ counters made out of plastic scintillator and buried at $2.5\,\mathrm{m}$ underground. 
The AMIGA muon detector accepts events up to $45^\circ$ of zenith angle.
Each muon counter is divided into 64 scintillator strips of
equal size, the three counters installed at each array position are equivalent to a single $30\,\mathrm{m^2}$ detector divided into $192$ bars. Muons are counted in time windows of $25\,\mathrm{ns}$, duration corresponding to the detector
dead time given by the width of the muon pulse due to the scintillator decay time.

We present a likelihood to reconstruct the muon lateral distribution function (LDF) with AMIGA. The new likelihood builds on two different methods we previously used. In the earlier one a likelihood only
valid for few muons in a detector was adopted~\cite{Supanitsky:2008dx}. This limitation reduced the number of events
that AMIGA could reconstruct successfully. To enlarge the reconstructed sample we later proposed an exact
likelihood~\cite{Ravignani:2014jza}. However in this second case the time resolution of the detector had to be neglected to obtain an analytic expression for the likelihood. The exact method only considers whether a bar has a signal during the whole
duration of the event. The likelihood introduced here combines the detector time resolution of the original method with
the extended dynamic range of the second one. 

The following section describes the likelihood, and section~\ref{sec:simulations} presents the simulations used
to evaluate it. In section \ref{sec:amiga} the new method is compared to the exact one. We finally conclude in section~\ref{sec:conclusions}.   
\newpage

\begin{figure} [!thp]
\centering
\setlength{\abovecaptionskip}{0pt}
\includegraphics[width=.5\textwidth]{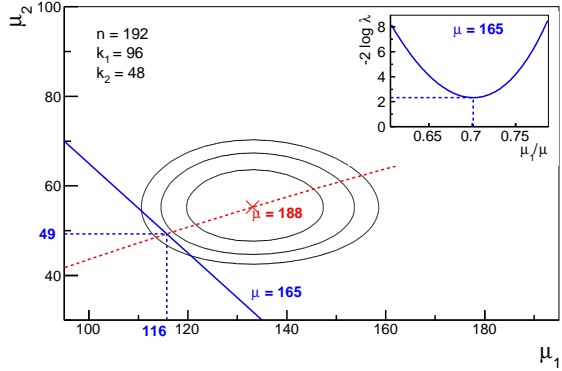}
\caption{Contour levels of $-2\,\log\lambda(\mu_1,\mu_2)$ for a signal spread over two time bins. In this example the detector is divided into 192 segments, the
first time bin has 96 bars \emph{on} and the second one 48. The red cross indicates the position of the likelihood minimum. Contour lines show increments of one from this minimum. The full blue line shows a cut at $\mu=165$. A local minimum along this cut is reached at $\mu_1=116$ and $\mu_2=49$. The red dotted line shows the minima location as function of $\mu$. {\bf Inset}: Function $-2\,\log\lambda(\mu_1,\mu_2)$ along the cut $\mu=165$. The profile likelihood value is the minimum of this function, in this case
$-2\,\log\lambda_p(\mu=165)\,=\,2.3$. \label{fig:like2b}}
\end{figure} 

\section{Profile likelihood}
\label{sec:likelihood}

To extend the likelihood to many time bins, one has to consider the time spread of the signal left by muons in the
detector. The total number of muons ($\mu$) is the integral of this signal for the whole event
duration. Within a time bin the number of muons ($\mu_i$) is the integral during the window duration. The sum of the $\mu_i$'s is $\mu$. 

As quoted in the introduction the AMIGA modules count muons in windows of $25\,\mathrm{ns}$ for each scintillator strip. Each time bin can be in two possible states, the state is \emph{on} if
a muon signal is detected during the bin duration or \emph{off} otherwise. The number of strips \emph{on} in each
time bin ($k_i$) is computed afterwards. For each time window the exact likelihood ($\mathcal{L}_i$)
from~\cite{Ravignani:2014jza} is used. Considering that the $k_i$'s are independent from each other, the full likelihood
($\mathcal{L}$) is the product of the likelihoods of each bin, 

\begin{equation}
  \mathcal{L}(\vec{\mu}) = \prod_{i=1} \mathcal{L}_i(\mu_i) = \prod_{i=1} 
  \binom{n}{k_i} \, e^{-\mu_i} \, \left( e^{\nicefrac{\mu_i}{n}} -1 \right)^{k_i},  
  \label{eq:likelihood}
\end{equation} 

\noindent where $i$ runs over the time bins, $n$ is the number of bars, and $\vec{\mu} = (\mu_1,\mu_2,\dotsc)$.
The likelihood is based on a model that considers the finite detector size and segmentation. However it does not account
for the signal contamination produced by a muon crossing two scintillator bars or the cross talk between PMT channels. 
These two effects are mitigated by the counting technique used in AMIGA~\cite{Wundheiler:2011zz}. The maximum likelihood estimator of 
the number of muons in each time window $\hat{\mu}_i$ is:  

\begin{equation}
\hat{\mu}_i = -n \, \log\left(1-\frac{k_i}{n}\right). 
\label{eq:muest}
\end{equation}

\begin{figure} [!thp]
\centering
\setlength{\abovecaptionskip}{0pt}
\includegraphics[width=.45\textwidth]{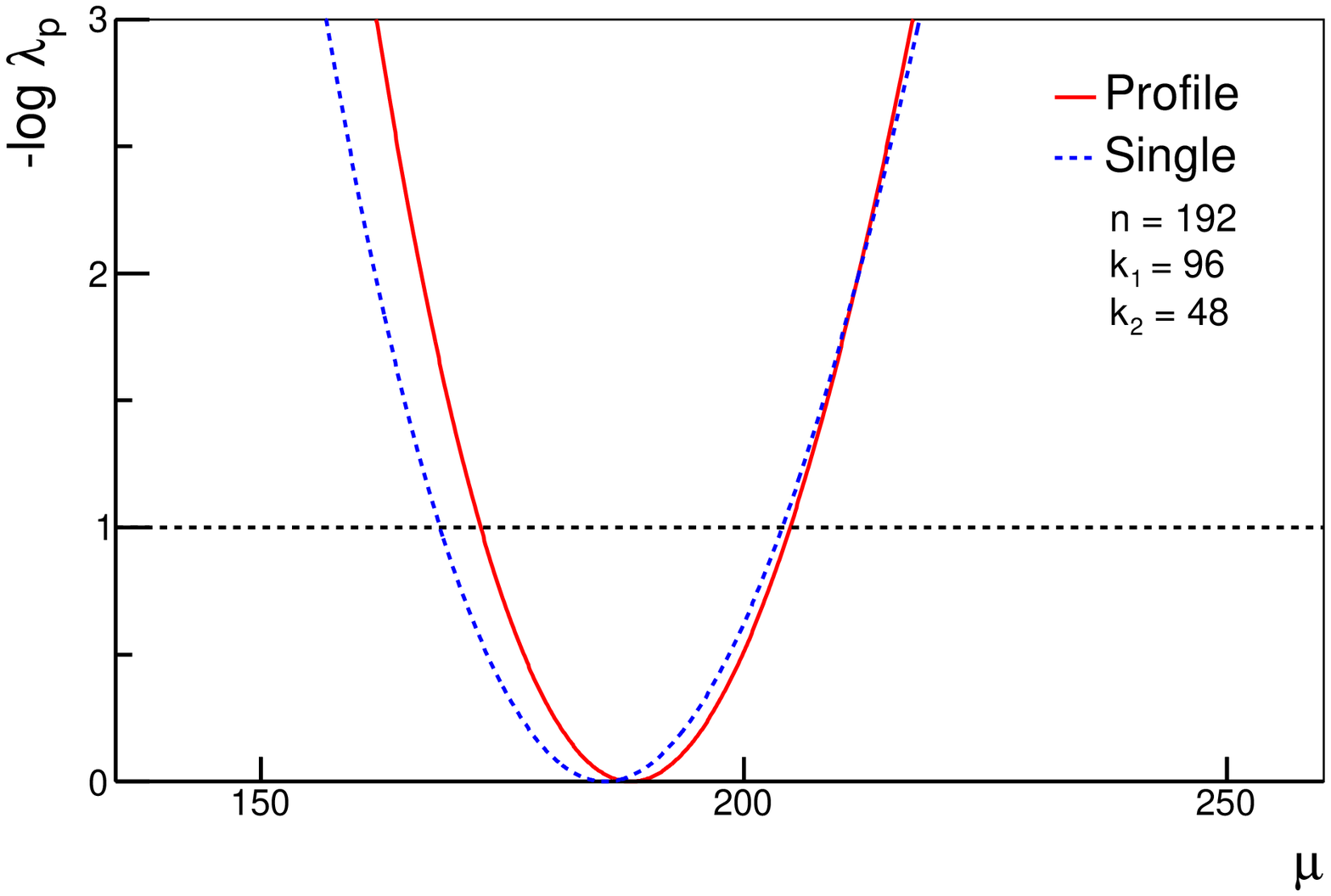}
\includegraphics[width=.45\textwidth]{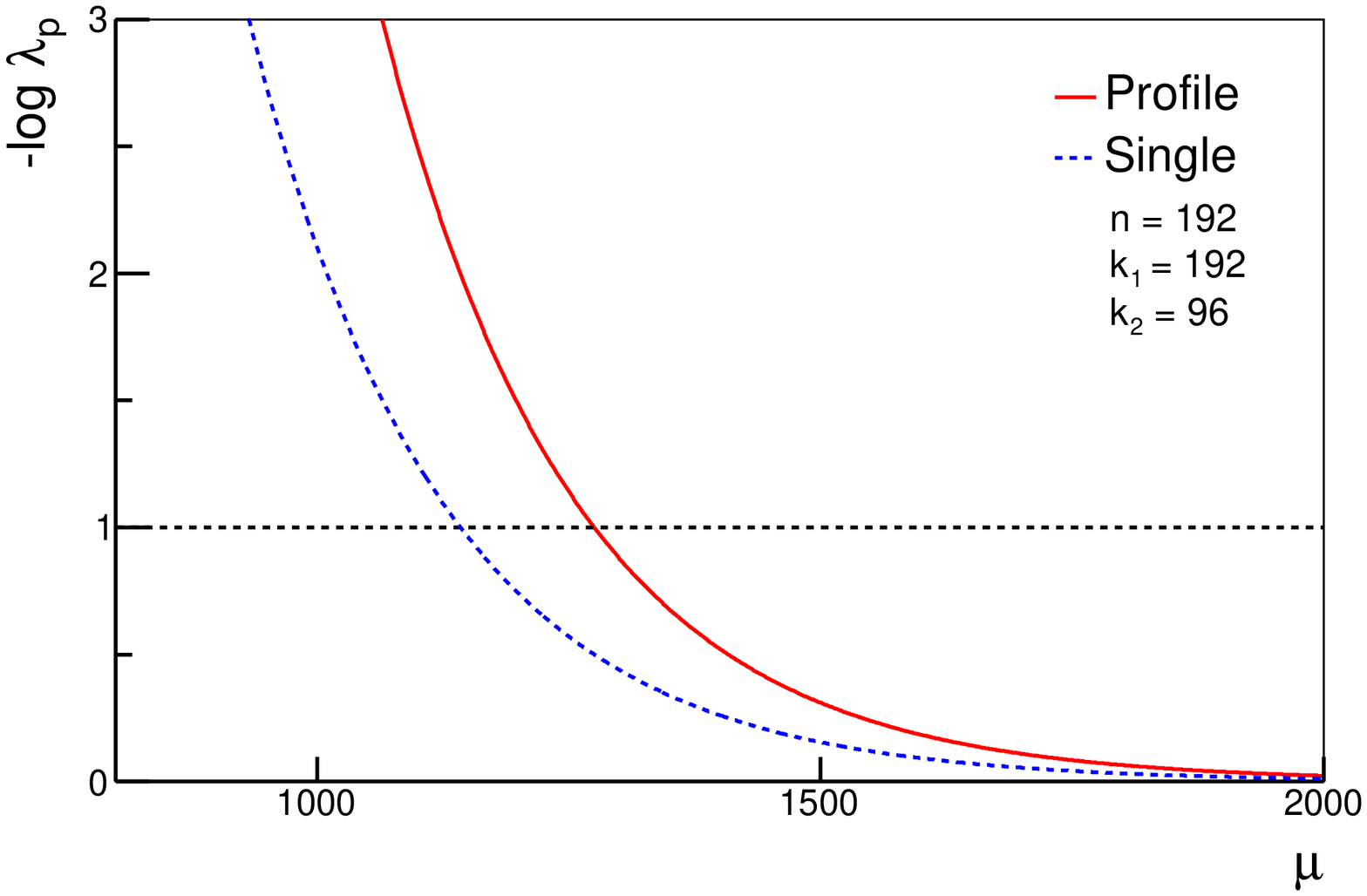}
\caption{Examples of the profile and single window likelihoods for a detector divided into $192$ segments. Left: Counter
with $96$ and $48$ segments \emph{on} in the first and second time bins respectively. For the single window likelihood
$119$ bars \emph{on} for the whole event duration were assumed. Right: Saturated detector with $192$ and $96$ bars
\emph{on} in the first and second time bins. \label{fig:like1}}
\end{figure}

The likelihood is maximised at a value $\mathcal{L}_{max}$ when ${\mu_i}=\hat{\mu}_i$ for all
windows. The maximum likelihood estimator of the number of muons in the detector is $\hat{\mu} =
\sum_{i=1} \hat{\mu}_i $. Instead of $\mathcal{L}$ we use the likelihood ratio
($\lambda(\vec{\mu})$) for the link it has to a $\chi^2$ distribution as will be presented later:   

\begin{equation}
\lambda(\vec{\mu}) = \frac{\mathcal{L}(\vec{\mu}) }{ \mathcal{L}_{max} }.
\label{eq:likelihooRatio}
\end{equation} 

The likelihood ratio is between 0 and 1, reaching the later at the maximum. The likelihood ratio
$\lambda(\vec{\mu})$ and $\mathcal{L}$ cannot be calculated because they depend, via the $\mu_i$'s, on the signal time
distribution which is not known. But this limitation can be avoided because only the number of muons $\mu$,
and not the signal timing, is required in the LDF fit. So we used instead an approximation called profile likelihood
($\lambda_p$)~\cite{Agashe:2014kda} that depends on $\mu$ only. In this approximation the maximum of $\lambda(\vec{\mu})$ is searched
varying the $\mu_i$'s with the restriction $\sum_{i=1} \mu_i = \mu$. The maximum value found is assigned to
$\lambda_p$:  

\begin{equation}
\lambda_p(\mu) = \max_{\sum\mu_i=\mu} \lambda(\vec{\mu}). 
\label{eq:profileLikelihood}
\end{equation} 

In figure~\ref{fig:like2b} the procedure to obtain the profile likelihood is illustrated using an example of a signal
spread over only two time windows. In practice for the reconstruction we used the function $f(\mu)=-2\,\log\lambda_p(\mu)$
which approximates a $\chi^2$ distribution~\cite{Agashe:2014kda} when there are many bars \emph{on} as happens in the
example. The function $f(\mu)$ is positive and drops to zero at $\hat{\mu}$. 
The function $f(\mu)$ of the presented example is shown in the left pane of figure~\ref{fig:like1}. In this plot we
compare the profile likelihood with the exact single window likelihood. Recurring to the analogy between $f(\mu)$ and a
$\chi^2$ distribution, approximate 1$\sigma$ confidence intervals are defined by the condition $f(\mu)=1$. The single
window likelihood has larger confidence intervals than the profile likelihood. This widening of the uncertainty equates
to a reduced detector resolution with the single window likelihood. The worsening of the detector response is expected
because the single window likelihood does not take advantage of the detector time resolution. 

Although counters saturate when
all bars are \emph{on}, the cases of the single window and profile likelihoods are different. While in the single
window likelihood the number of bars \emph{on} in the whole event matters, with the profile likelihood only the
situation within a single time bin applies. For a signal spread over many time bins saturation starts at lower signals
in the single window than in the profile likelihood. The two likelihoods are displayed in the right panel of
figure~\ref{fig:like1} for an example of a saturated detector. In both cases the saturated likelihood sets a lower bound to the values allowed for $\mu$ in the LDF fit. The profile likelihood imposes a more
stringent limit than the single window method.  

\begin{figure}[thp]
\centering
\setlength{\abovecaptionskip}{0pt}
\includegraphics[width=.45\textwidth]{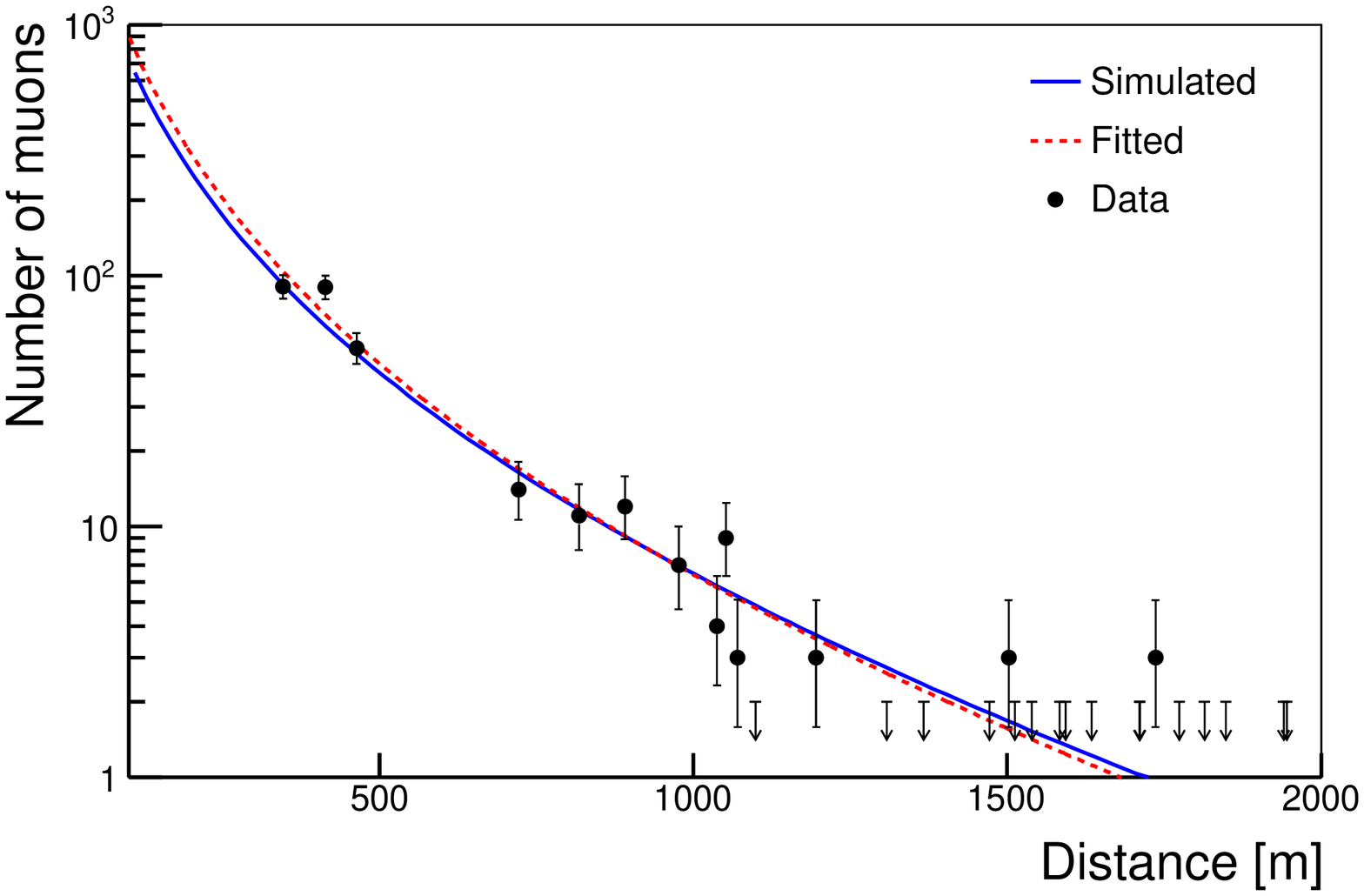}
\includegraphics[width=.45\textwidth]{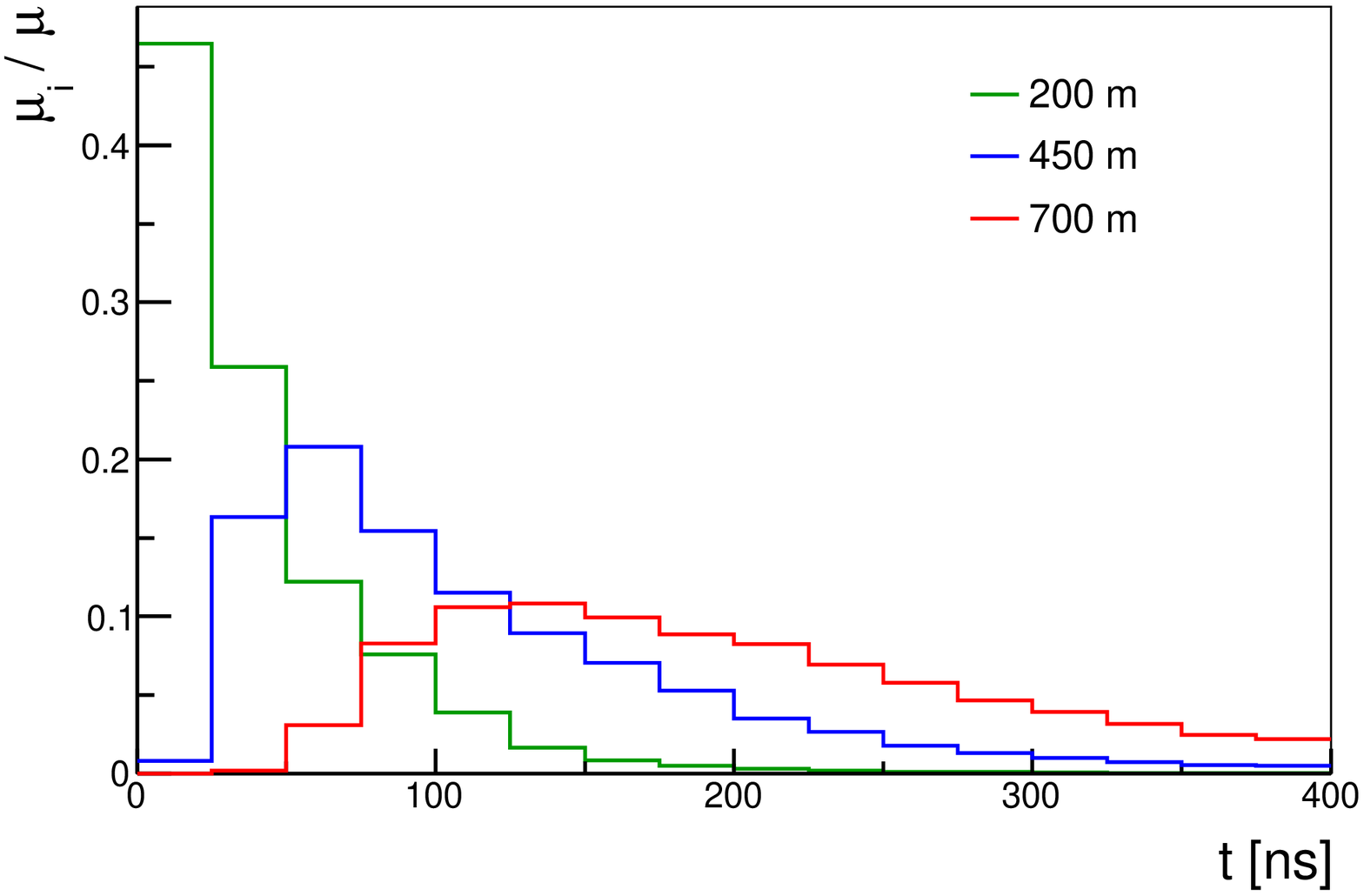}
\caption{Left panel: True average muon lateral distribution function obtained from simulations of an iron primary with
energy $E=1\,\mathrm{EeV}$ and zenith angle $\theta=30^{\circ}$. Each point corresponds to the number of estimated muons
in a counter $\hat{\mu}$. The untriggered counters are represented with a down arrow. A fit of
simulated data using the profile reconstruction is also shown. Right panel: Average time distributions for three
different distances to the shower axis. The histograms show the fractions of muons in each $25\,\mathrm{ns}$ time bin with
respect to the total number of muons. At longer distances the muons arrive more spread. \label{fig:sim}}
\end{figure} 

\section{Simulations}
\label{sec:simulations}

To test the performance of the profile and single window likelihoods we simulated air showers with CORSIKA
v7.3700~\cite{Knapp:1998ra} using the high energy hadronic model EPOS-LHC~\cite{Pierog:2013ria}. Proton and iron
primaries were run in the energy interval $\log_{10}(E/\textrm{eV}) \in [17.5, 19]$ in steps of
$\Delta\log_{10}(E/\textrm{eV}) = 0.25$ for the zenith angles $\theta = 0^\circ$, $30^\circ$, and $45^\circ$. We used a
thinning algorithm with an optimal statistical thinning of $10^{-6}$. Twenty proton and fifteen iron showers were
simulated for each energy and zenith angle combination. From these showers an average LDF was fitted using a
KASCADE-Grande like function~\cite{Apel:2010zz}. An average distribution of the muon arrival times was also obtained.
Figure~\ref{fig:sim} shows the average LDF and time distribution of $1\,\mathrm{EeV}$ iron showers arriving
at $\theta=30^{\circ}$.   

For each energy and zenith angle combination we ran 10000 shower reconstructions varying the azimuth angle and the position of the
impact point on the ground. In each reconstruction the distances of counters to the shower axis were calculated.
The average number of muons simulated in each counter ($\mu$) was taken from the LDF.
Using $\mu$ as a parameter the actual number of muons was sampled from a Poisson distribution. 
A detector was considered \emph{untriggered} if it received two or fewer muons. 
The arrival time of each muon was obtained from the corresponding time distribution. The number of
muons $\mu$ as function of the shower axis distance was fitted to the simulated data with another KASCADE-Grande like function, 

\begin{equation}
\label{LDF}
\mu(r)=A_\mu \left( \frac{r}{r_1} \right)^{-\alpha} \! \left( 1+\frac{r}{r_1} \right)^{-\beta}%
\! \left( 1+\left( \frac{r}{10 \, r_1}\right)^2 \right)^{-\gamma} \! \! \! \!  , 
\end{equation}

\noindent where $r$ is the distance to the shower axis, $\alpha = 0.75$, $r_1 = 320 \,\mathrm{m}$, and $\gamma=2.95$. The
normalisation $A_\mu$ and the slope $\beta$ were adjusted in the LDF fit by maximising a likelihood. Both the single
window and profile likelihoods were used. For untriggered counters we used a Poisson likelihood setting an upper limit
to the number of allowed muons in the LDF fit as in Ref.~\cite{Supanitsky:2008dx}. The left panel of figure~\ref{fig:sim}
shows the fit of a simulated shower using the profile likelihood.

\section{Reconstruction performance}
\label{sec:amiga}

\begin{figure} [thp]
\centering
\setlength{\abovecaptionskip}{0pt}
\includegraphics[width=.45\textwidth]{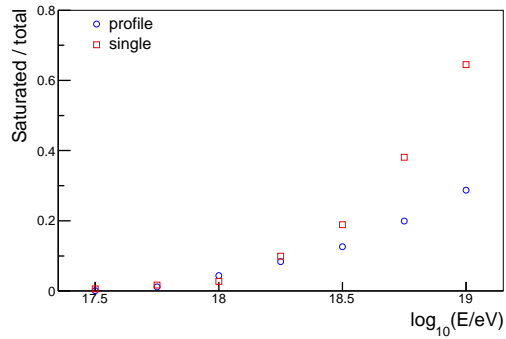}
\caption{Number of saturated events with respect to the total number of simulated events as function of energy for the
profile and single window reconstructions. An iron primary at $\theta = 30^\circ$ was used in this example.
\label{fig:satAmiga}} \end{figure}

In this section we compare the performance of the LDFs reconstructed with the profile and the single window
likelihoods. The evaluation of the LDF at a reference distance is the established way to derive an estimator of
the shower size~\cite{Newton:2006wy}. In this section we find a reference distance of $450\,\mathrm{m}$ for AMIGA and calculate the
bias and variance of the corresponding shower size parameter. From the set of events used to find them we excluded those that have a saturated
counter because their reconstructed LDF is
biased~\cite{Ravignani:2014jza}. As already mentioned in section~\ref{sec:likelihood} saturation occurs before with the single
window than with the profile likelihood. The number of saturated events for the two likelihoods are displayed in
figure~\ref{fig:satAmiga} for an iron primary at $\theta = 30^\circ$. As expected there are more saturated events with
the single window than with the profile likelihood. The analysis of the single window reconstruction was cut at
$E=10^{18.75}\,\mathrm{eV}$ because more than $60\%$ of the events are saturated at this energy.

\begin{figure} [thp]
\centering
\setlength{\abovecaptionskip}{0pt}
\includegraphics[width=.45\textwidth]{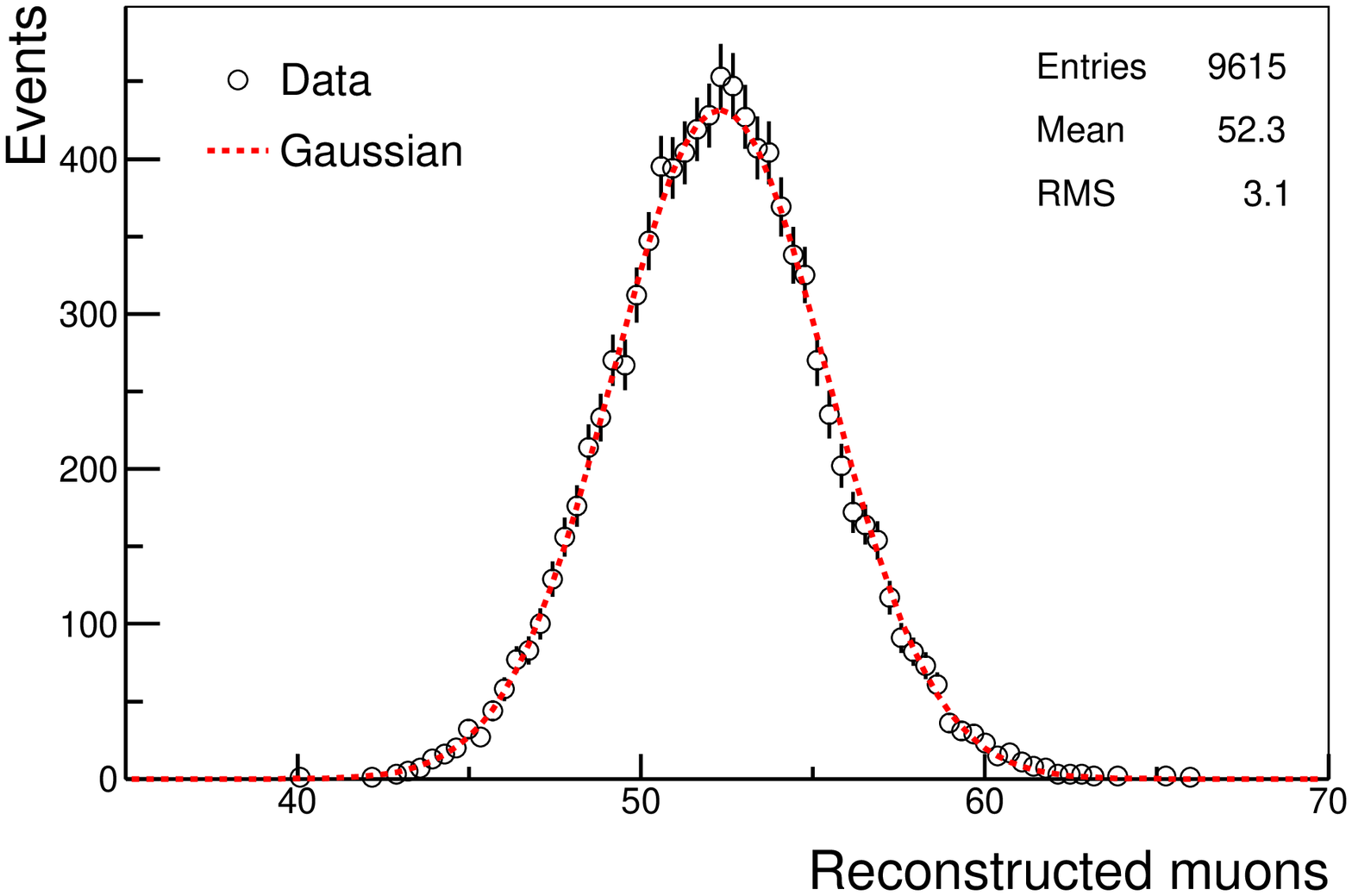}
\includegraphics[width=.45\textwidth]{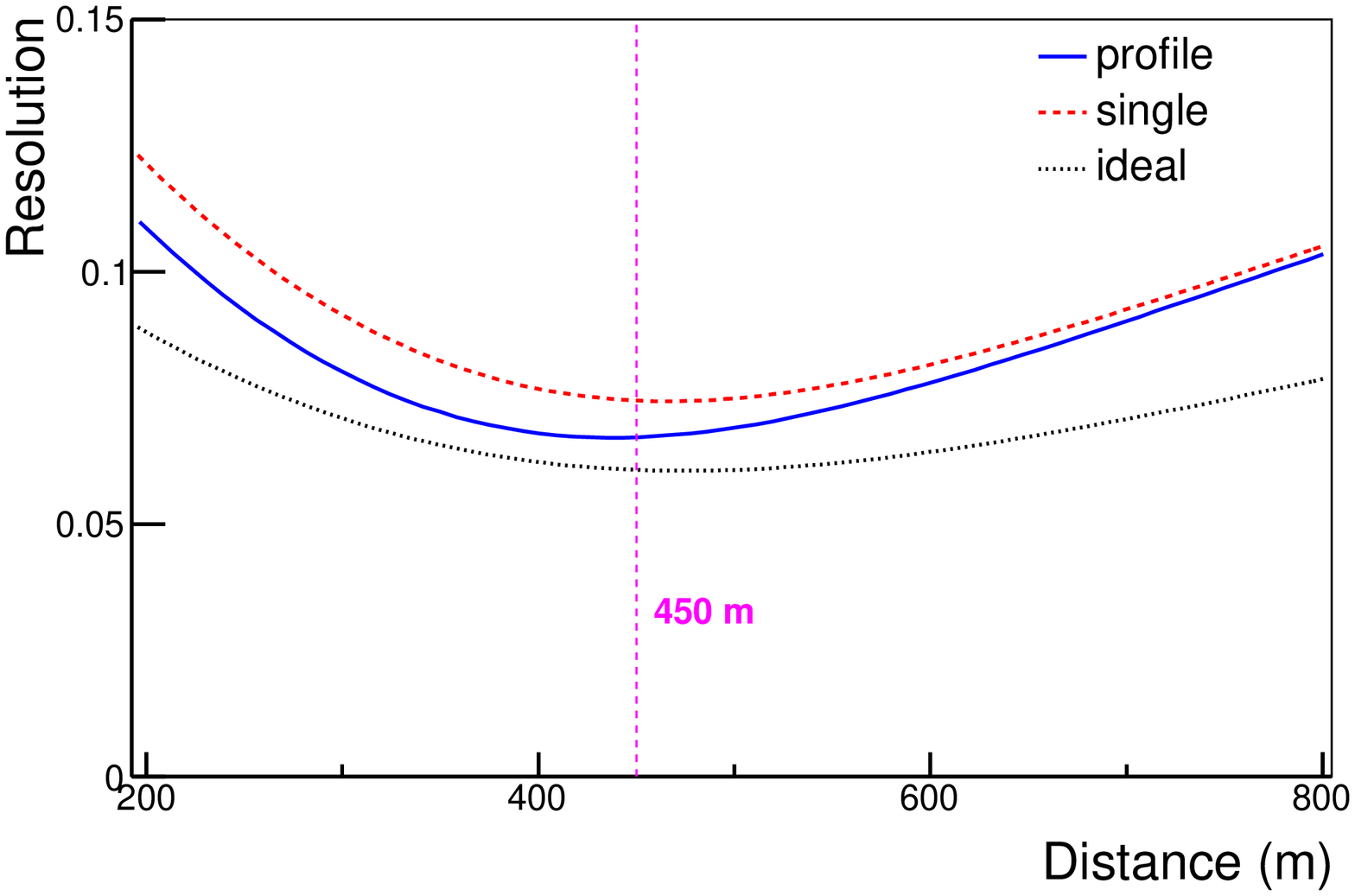}
\caption{Left: Distribution of the number of muons reconstructed at $450\,\mathrm{m}$ with the profile likelihood. Simulations of $1\,\mathrm{EeV}$ iron showers at $\theta = 30^\circ$ were used. In this example the detector relative standard deviation is $\varepsilon=6\%$. Right: AMIGA resolution as function of the distance to the shower core ($\varepsilon(r)$). We used proton and iron showers at all simulated energies and zenith angles. The profile, single window, and ideal cases reach a minimum around $450\,\mathrm{m}$. \label{fig:rminAmiga}}
\end{figure}

For each event the reconstructed LDF was evaluated at the reference distance of $450\,\mathrm{m}$. 
The $\mu(450)$ obtained in this way fluctuates across reconstructions of the same shower given the uncertainty in
the number of muons measured by the counters.  
An histogram of the $\mu(450)$ reconstructed using the profile likelihood is shown
in figure~\ref{fig:rminAmiga} for $1\,\mathrm{EeV}$ iron showers at $\theta = 30^\circ$. 
A Gaussian function parametrised with the histogram mean and standard deviation
$(\sigma(450))$ is displayed. The $\mu(450)$ is unbiased, its mean matches the true value taken
from the average LDF used as the reconstruction input. The standard deviation $\sigma(450)$ measures the fluctuations of
the reconstructed value around the true parameter. The relative standard deviation ($\varepsilon$) is defined as the standard deviation $\sigma$ relative to the number of muons $\mu$. 

\begin{figure} [thp]
\centering
\setlength{\abovecaptionskip}{0pt}
\includegraphics[width=.45\textwidth]{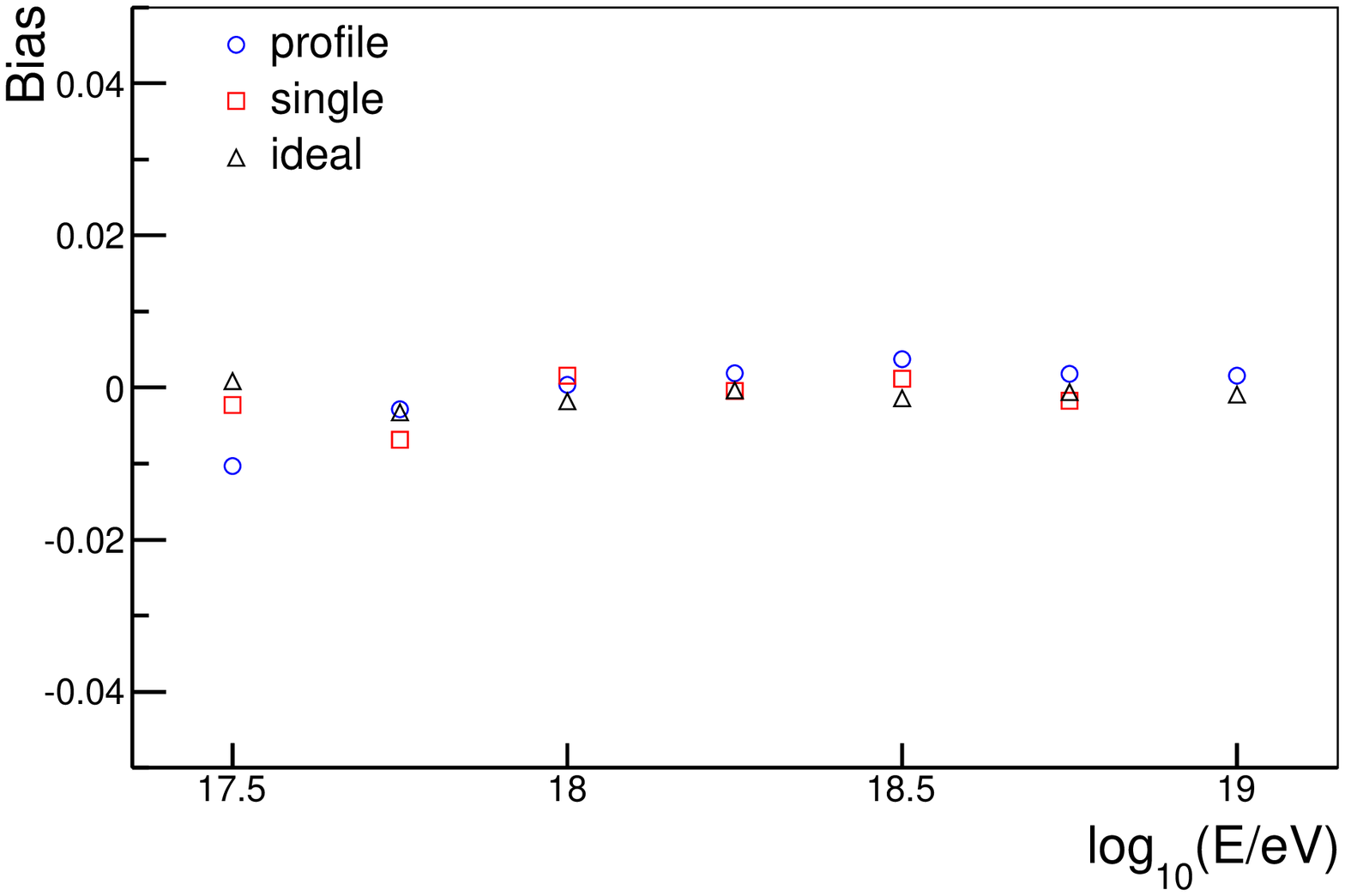}
\includegraphics[width=.45\textwidth]{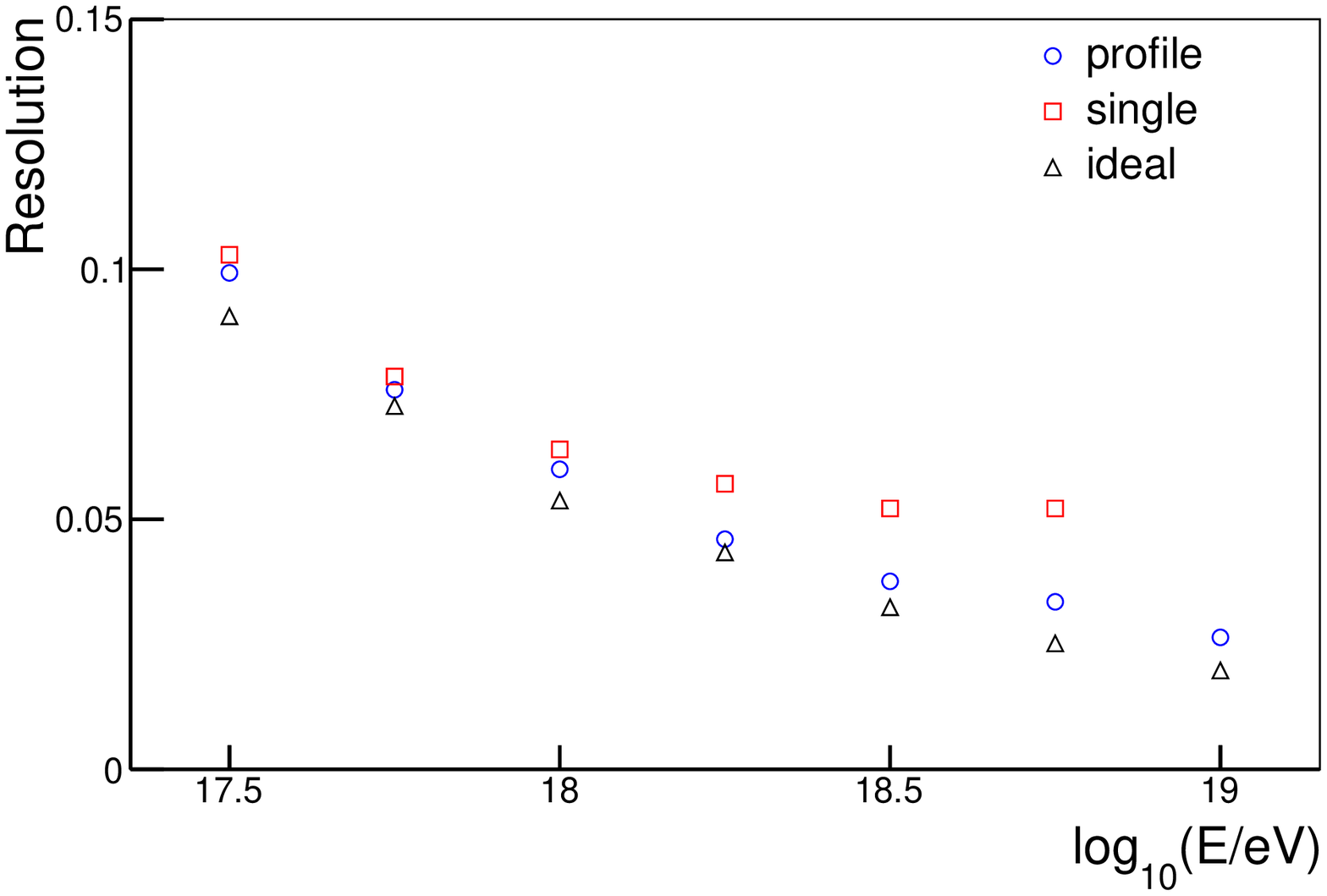}
\caption{Relative bias (left) and standard deviation (right) of the $\mu(450)$ reconstructed with the single window and profile likelihoods and with an ideal counter. The data correspond to iron showers at $\theta = 30^\circ$. \label{fig:biasSigmaAmiga}}
\end{figure} 

The standard deviation of the reconstructed LDF depends on the distance to the core at which this function is evaluated.
We selected a distance of $450\,\mathrm{m}$ to minimise the LDF fluctuations. 
To find the optimal distance we computed the relative standard deviation of each shower type and added them in
quadrature to obtain a global $\varepsilon(r)$. 
The right panel of figure~\ref{fig:rminAmiga} shows $\varepsilon(r)$ for
the single window and the profile likelihoods. The case of an ideal counter that has infinite segments is also included. The
ideal counter, already introduced in~\cite{Ravignani:2014jza}, sets a lower bound to the resolution achievable with a
detector of the AMIGA size. 
Because the counter uncertainties using the single window are
higher than with the profile likelihood the corresponding fluctuations in $\mu(450)$ also are.
The optimal distance is close to $450\,\mathrm{m}$ in the single window,
profile, and ideal counter reconstructions. 

We compared the likelihoods based on the bias and the standard deviation of the $\mu(450)$ reconstructed with each of them. We recall that only unsaturated events were used for these benchmarks. 
In figure~\ref{fig:biasSigmaAmiga} the relative bias and standard deviation of $\mu(450)$ are shown as function of energy for iron showers at $\theta = 30^\circ$. 
The three reconstructions have biases below $2\%$ in all the energy range. The relative standard deviation of the three methods is also similar up to $1\,\mathrm{EeV}$, but at higher energies the profile reconstruction has a better resolution than the single window one. 
With the single window likelihood the resolution of the counters deteriorates as muons start to accumulate in the same scintillator strip. 
The effect is more noticeable at high energy, when there are more muons and therefore they pile up more. 
On the other hand for the profile likelihood muons are distributed over many windows so there are fewer muons per time bin. 
The resolution achieved with the profile likelihood is similar to the best case set by the ideal counter in the considered energy range.
The same results about saturation, bias, and standard deviation stand for all simulated showers.

\section{Conclusions}
\label{sec:conclusions}

We introduced a likelihood to reconstruct the muon lateral distribution function with an array of segmented counters like AMIGA.
The new method combines the detector time resolution used in the original AMIGA reconstruction with the exact probability
function developed afterwards. 
We showed that the profile likelihood improves the reconstruction with respect to the exact alternative in two aspects. 
Firstly, by raising the saturation limit of muon counters, more events useful for science analyses can be reconstructed. 
Secondly, the profile likelihood improves the detector resolution allowing for a more powerful discrimination between different primary masses.

\end{document}